\documentclass[epj]{webofc}
\usepackage[utf8]{inputenc}
\usepackage[varg]{txfonts}
\usepackage{eucal}
\usepackage{booktabs}
\usepackage{xcolor}
\usepackage{wrapfig}
\definecolor{darkred}{rgb}{0.4,0.0,0.0}
\definecolor{darkgreen}{rgb}{0.0,0.4,0.0}
\definecolor{darkblue}{rgb}{0.0,0.0,0.4}
\usepackage[bookmarks,linktocpage,colorlinks,
    linkcolor = darkred,
    urlcolor  = darkblue,
    citecolor = darkgreen]{hyperref}
\wocname{EPJ Web of Conferences}
\woctitle{Lattice2017}

\def\rmd{{\rm d}}

\def\rme{{\rm e}}
\def\rmO{{\rm O}}

\def\proof{\noindent{\sl Proof:}\kern0.6em}

\def\sfrac#1#2{\hbox{\normalsize$\frac{#1}{#2}$}}
\def\dual{\mathstrut^*\kern-0.1em}

\def\slash#1{\setbox2=\hbox{$\displaystyle#1$}%
             \setbox3=\hbox{$\displaystyle/$}%
             #1\kern-0.8\wd2/\kern-1.0\wd3\kern0.8\wd2\kern0.5pt}
\def\lvec#1{\setbox0=\hbox{$#1$}
    \setbox1=\hbox{$\scriptstyle\leftarrow$}
    #1\kern-\wd0\smash{
    \raise\ht0\hbox{$\raise1pt\hbox{$\scriptstyle\leftarrow$}$}}
    \kern-\wd1\kern\wd0}
\def\rvec#1{\setbox0=\hbox{$#1$}
    \setbox1=\hbox{$\scriptstyle\rightarrow$}
    #1\kern-\wd0\smash{
    \raise\ht0\hbox{$\raise1pt\hbox{$\scriptstyle\rightarrow$}$}}
    \kern-\wd1\kern\wd0}
\def\cvec#1{\kern-0.5pt\vec{\kern0.5pt #1}}
\def\boxit#1{\vbox{\hrule height2pt\hbox{\vrule width2pt
    \kern10pt\vbox{\kern10pt#1\kern10pt}\kern10pt\vrule width2pt}
    \hrule height2pt}}

\def\llangle{\langle\!\langle}
\def\rrangle{\rangle\!\rangle}

\def\nab#1{{\nabla\kern-1.5pt_{#1}\kern1.0pt}}
\def\nabstar#1{\nabla\kern0.5pt\smash{\raise5.0pt\hbox{$\ast$}}
               \kern-7.5pt_{#1}\kern2.0pt}

\def\drvstar#1{\partial\kern1.0pt\smash{\raise4.5pt\hbox{$\ast$}}
               \kern-6.5pt_{#1}\kern2pt}

\def\MSbar{\overline{\rm MS\kern-0.5pt}\kern0.5pt}

\def\diracstar#1#2{
    \setbox0=\hbox{$\gamma$}\setbox1=\hbox{$\gamma_{#1}$}
    \gamma_{#1}\kern-\wd1\kern\wd0
    \smash{\raise4.5pt\hbox{$\scriptstyle#2$}}}

\def\obs{{\mathcal O}}
\def\obsb{\overline{\vphantom{(}\obs}}

\def\eps{\epsilon}
\def\epsm{\eps_{\rm max}}
\def\mpi{m_{\pi}}
\def\Pb{P_{\Lambda}}
\def\Db{D_{\Lambda}}

\begin{document}
\selectlanguage{english}

\rightline{CERN-TH-2017-171}

\title{Stochastic locality and master-field simulations of very large\\
lattices%
\footnote{Talk given at the 35th International Symposium on
Lattice Field Theory, 18-24 June 2017, Granada, Spain.}}
\author{\firstname{Martin} \lastname{L\"uscher}\inst{1,2}
}
\institute{
CERN, Theoretical Physics Department, 1211 Geneva 23, Switzerland,
\and
Albert Einstein Center for Fundamental Physics,
Sidlerstrasse 5, 3012 Bern, Switzerland
}
\abstract{
In lattice QCD and other field theories with a mass gap,
the field variables in distant regions of a physically large lattice are only
weakly correlated.
Accurate stochastic estimates of the
expectation values of local observables
may therefore be obtained from a single representative field.
Such master-field simulations potentially allow very large
lattices to be simulated, but require various conceptual and technical issues
to be addressed. In this talk, an introduction to the subject is provided
and some encouraging results of master-field simulations of the SU(3) gauge
theory are reported.
}

\maketitle

\section{Introduction}\label{sect1}

Numerical simulations of Euclidean lattice field theories usually proceed by
generating an ensemble of representative fields through a Markov process.
The ensemble averages of the observables of interest then
provide stochastic estimates of their field-theoretical expectation values.

In lattice QCD,
the field variables in distant regions of a physically
large lattice fluctuate
largely independently, a property that may be referred to as
``stochastic locality''. Moreover, assuming periodic
boundary conditions, their distribution is
everywhere the same. The translation averages
\begin{equation}
  \llangle\obs(x)\rrangle=\frac{1}{V}\sum_z\obs(x+z)
  \label{TransAvg1}
\end{equation}
of local observables $\obs(x)$
(where $V$ denotes the number of lattice points and $z$ runs over
the lattice) are thus expected to coincide with their field-theoretical
expectation values,
\begin{equation}
  \llangle\obs(x)\rrangle=\langle\obs(x)\rangle+\rmO(V^{-1/2}),
  \label{TransAvg2}
\end{equation}
up to statistical errors of order $V^{-1/2}$.
These formulae also apply in the case of
Wilson loops, products of local fields and other non-local observables,
provided their localization range is much smaller than the
linear extent of the lattice.
On very large lattices, accurate results for the expectation values
of all these observables may therefore conceivably be obtained
from a single representative field.

Such master-field simulations need not require astronomical computer
resources. The computer time required for the generation
of $256$ statistically independent field configurations on an $L^4$ lattice
or of a single field on a $(4L)^4$ lattice is roughly the same.
Master-field simulations of QCD on lattices with $256^4$ points and
physical extent in the range from $12$ to $25$ fm, for example,
may thus very well be practically feasible at present
and would obviously be very interesting.
In this talk, various technical aspects of this new type of simulations
are addressed and some first numerical studies are presented for
illustration.

\section{Statistical error estimation}\label{sect2}

The statistical errors of ensemble averages of the observables
of interest are usually estimated from the variances
and statistical correlations of the measured values of the observables
\cite{UlliError}.
In master-field simulations,
where the generated ensemble of fields contains
a single field or at most a
few fields on a physically large lattice, the error estimation
must proceed in a different way. In particular,
the correlations of the field variables in space
must be taken into account.

\subsection{Single-field error formula}\label{subsect21}

Let $\obs(x)$ be an observable localized in
a region centered at the point $x$. For the derivation of
the error formula it is helpful to imagine
that a long simulation has been performed.
The translation averages $\llangle\obs(x)\rrangle$
of the observable calculated in the course of the
simulation are then distributed according to their field-theoretical
distribution. In the limit of an infinitely long run,
their average thus coincides with $\langle \obs(x)\rangle$
and their mean square deviation from the average is given by
\begin{equation}
  \bigl\langle\{\llangle\obs(x)\rrangle-\langle\obs(x)\rangle\}^2\bigr\rangle
   =
  \frac{1}{V}\sum_y\langle\obs(y)\obs(0)\rangle_{\rm c}.
  \label{StatErr1}
\end{equation}
In this equation, use has
been made of the periodic boundary conditions
and the translation invariance of
the two-point correlation function of the observable.
The subscript ``c'' indicates that
the correlation function is the connected one.

Equation (\ref{StatErr1}) provides an exact expression for the average
mean square deviation of the translation average $\llangle\obs(x)\rrangle$
from the field-theoretic expectation value $\langle\obs(x)\rangle$
in a master-field simulation. Since the correlation function in this
expression falls off exponentially at large distances, the formula
shows that the statistical errors in these simulations do in fact
decrease proportionally to $V^{-1/2}$ as $V\to\infty$. Moreover,
the proportionality constant is seen to be determined by the
two-point correlation function of the observable considered and
thus depends on the localization range of the latter and the
relevant correlation lengths.

As it stands, Eq.~(\ref{StatErr1}) is not particularly useful,
since the correlation function on the right is usually not known.
One may, however, sum the correlation function up to some radius $R$,
\begin{equation}
  \bigl\langle\{\llangle\obs(x)\rrangle-\langle\obs(x)\rangle\}^2\bigr\rangle
   =
   \frac{1}{V}\Bigl\{
     \sum_{|y|\leq R}\langle\obs(y)\obs(0)\rangle_{\rm c}
     +\rmO(\rme^{-mR})\Bigr\},
   \label{StatErr2}
\end{equation}
at the expense of an exponentially small systematic error, $m$
being some mass that characterizes the decay of the correlation function
at large distances.
The replacement of the correlation function through
the corresponding translation average then leads to a formula
\begin{equation}
  \bigl\langle\{\llangle\obs(x)\rrangle-\langle\obs(x)\rangle\}^2\bigr\rangle
   =
   \frac{1}{V}\Bigl\{
     \sum_{|y|\leq R}\llangle\obs(y)\obs(0)\rrangle_{\rm c}
     +\rmO(\rme^{-mR})+\rmO(V^{-1/2})\Bigr\}
  \label{StatErr3}
\end{equation}
suitable for numerical evaluation.
The last term on the
right of this equation, i.e.~the statistical ``error of the error'',
could be expressed through the four-point function of $\obs(x)$ in a
way similar to the error itself. Clearly,
Eq.~(\ref{StatErr3}) assumes the existence of a window of
values of $R$, where both the systematic error and the statistical error
of the error are negligible.

\subsection{Several master fields}\label{subsect22}

In the case of a (usually small) ensemble $\{U_1,\ldots,U_n\}$ of master fields,
generated by letting the simulation algorithm
run for some length of time, one first calculates the ensemble average
\begin{equation}
  \obsb(x)=\frac{1}{n}\sum_{k=1}^n\left.\obs(x)\right|_{U=U_k}
  \label{StatErr4}
\end{equation}
of the observable $\obs(x)$. The
translation average $\llangle\obsb(x)\rrangle$ then provides
an estimate of the expectation value $\langle\obs(x)\rangle$ with
variance
\begin{equation}
  \bigl\langle\{\llangle\obsb(x)\rrangle-\langle\obs(x)\rangle\}^2\bigr\rangle
   =
  \frac{1}{V}\Bigl\{
  \sum_{|y|\leq R}\llangle\obsb(y)\obsb(0)\rrangle_{\rm c}+\ldots\,
  \Bigr\},
  \label{StatErr5}
\end{equation}
provided the autocorrelation functions of the observable
are translation invariant and decay rapidly at large distances in space.
These technical requirements are usually satisfied and certainly so,
if the simulation algorithm is in the universality class of the (position-space)
Langevin equation.

Clearly, if the fields $U_1,\ldots,U_n$ are statistically independent,
the variance (\ref{StatErr5}) is exactly $n$ times smaller than
the variance in the case of a single
master field.
Equation (\ref{StatErr5}) however remains valid in
presence of statistical correlations and these need not be determined,
i.e.~they are automatically taken into account and
only the terms represented by
the ellipsis depend on them.

\subsection{Selected remarks}\label{subsect23}

\subsubsection{Application of the fast Fourier transform (FFT)}

The correlation function on the right of the error formula (\ref{StatErr3}),
\begin{equation}
  \sum_{|y|\leq R}\llangle\obs(y)\obs(0)\rrangle_{\rm c}
  =
  \frac{1}{V}\sum_{|y|\leq R}\sum_z
  \bigl(\obs(y+z)-\llangle\obs(y)\rrangle\bigr)
  \bigl(\obs(z)-\llangle\obs(0)\rrangle\bigr),
  \label{StatErr6}
\end{equation}
may be evaluated in two steps. Assuming $\obs(z)$ is known at all lattice
points, the sum over $z$ can first be performed
for all $y$ simultaneously using the FFT.
The sum over $y$ may then be evaluated for all $R$ in the chosen
range of values. As $V\to\infty$, the numerical effort required
for the calculation of the sums
in Eq.~(\ref{StatErr6}) thus scales like $V\ln V$.

\subsubsection{``Dilute data''}\label{subsubsect232}

In practice an observable $\obs(z)$ may only be known
on a sublattice of points $z$.
The translation averages are, in these cases, taken over
the sublattice and the sums in the error formulae then run over the
sublattice too. Depending on the correlations of the observable
in space, the restriction
to the sublattice may or may not have a significant influence
on the statistical errors.

\subsubsection{Error correlations and propagation}

The error formulae derived in this section
straightforwardly generalize to the case
of the statistical correlations of two or more primary
observables.
Bootstrap methods then provide
a computationally efficient and convenient way to propagate the statistical
errors to any secondary observables.

\section{Sample calculations}\label{sect3}

The master-field simulations of the SU(3) gauge theory
reported in this section mainly serve to demonstrate that the error
formulae derived in the previous section work out in practice.
In all cases, the Wilson plaquette action \cite{Wilson} was used and
the gauge fields were generated with the SMD algorithm
(see sect.~\ref{sect4}). Lattices of size up
to $192^4$, with spacing $a$ equal to $0.10$ or $0.05$ fm, were
simulated. The simulations of the largest lattices occupied
$64$ nodes ($1536$ cores, $8.2$ TB total
memory) of the FinisTerrae II machine at
CESGA for about 10 days\footnote%
{See {\tt https://www.cesga.es/en/infraestructuras/computacion/FinisTerrae2}
for a description of the machine.}.

\subsection{Reference flow time}\label{subsect31}

Among the most easily accessible renormalized quantities in
the pure gauge theory are the Yang-Mills action
density $E(x)$ at gradient-flow time $t>0$ and the reference scale $t_0$
that derives from the latter \cite{WilsonFlow,RenFlow}.
The action density $E(x)$ is a primary observable localized
in a ball centered at $x$ with radius roughly equal to $\sqrt{8t}$
(at flow times near $t_0$, this is about $0.5$ fm in physical units).

\begin{figure}[t] 
  \centering
  \includegraphics[width=6.0cm,clip]{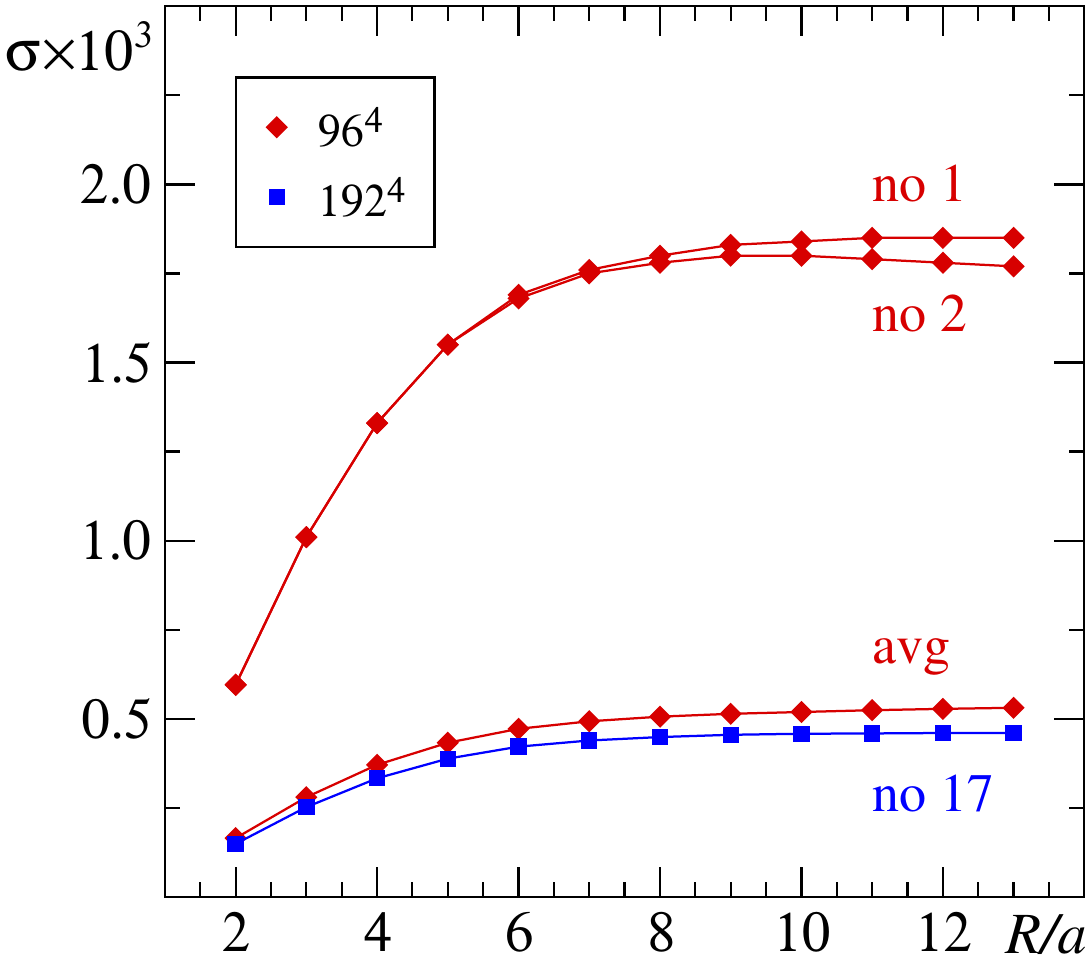}
  \caption{Relative statistical error $\sigma$
  of $\llangle E(x)\rrangle$ at gradient-flow time $t\simeq t_0$,
  as determined by the first term on the right of the error
  formulae (\ref{StatErr3}) and (\ref{StatErr5}), respectively.
  The lines connecting the data points are drawn to
  guide the eyes. Except for the data labeled ``avg'', which are
  obtained from an ensemble of $13$ fields, the data are from different
  single fields (the numbering of the fields
  corresponds to the one in Fig.~\ref{fig2}).}
  \label{fig1}
\end{figure}

The calculations of the action density and of the reference flow time
reported here were performed on lattices of size $96^4$ and $192^4$
with spacing $a=0.10$ fm. On these lattices and at $t\simeq t_0$,
the statistical errors of $\llangle E(x)\rrangle$
safely reach a plateau as a function of the summation
radius $R$ used in the error formulae (see Fig.~\ref{fig1}).
There is a
small difference in the plateau height obtained in the case
of the fields no~1 and 2, which gives an indication of the
size of the error of the error on the $96^4$ lattice.
Averaging the values of $E(x)$ measured on $n=13$
weakly correlated configurations
has the expected effect of lowering the statistical error
by the factor $n^{-1/2}$
(points labeled ``avg'' in Fig.~\ref{fig1}).
Even smaller errors are obtained when the lattice size is
increased from $96^4$ to $192^4$. The (relative)
error of the error then also goes down by about a factor $4$.

\begin{figure}[t] 
  \centering
  \includegraphics[width=10.3cm,clip]{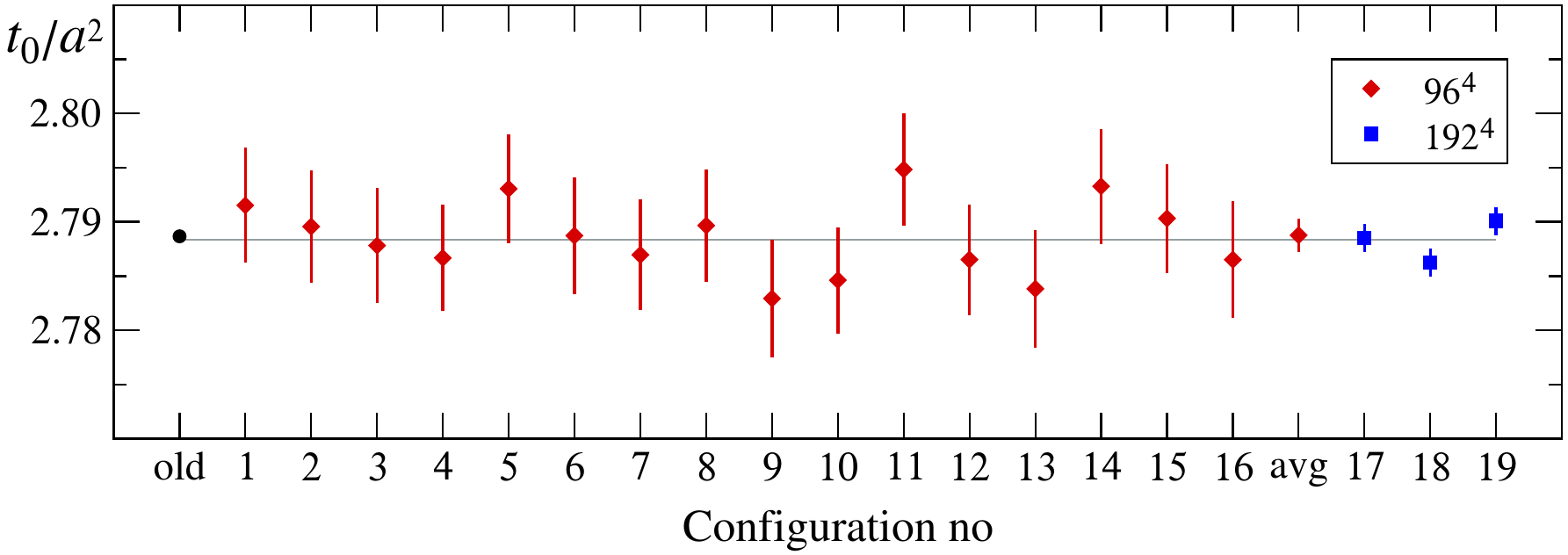}
  \caption{Values of the reference flow time $t_0$ obtained in
  master-field simulations. The point labeled
  ``old'' is from a previous traditional simulation \cite{CeEtAlTop}
  and the one labeled ``avg'' from an ensemble of $13$ fields on
  the $96^4$ lattice. A least-squares fit of
  the single-field results was used to produce the horizontal
  line.
  }
  \label{fig2}
\end{figure}

As shown by Fig.~\ref{fig2}, the results for the reference flow time $t_0$
obtained in the master-field simulations reported here are
statistically compatible among themselves and with an earlier
traditional calculation. Given the estimated statistical uncertainties,
the scattering of the points is certainly plausible and thus
supports the conclusion that the errors are correctly
estimated.

\subsection{Topological susceptibility}\label{subsect32}

Following ref.~\cite{WilsonFlow}, the topological susceptibility
is defined by
\begin{equation}
  \chi_t=\sum_y
  \langle q(y)q(0)\rangle,
  \label{chit1}
\end{equation}
where $q(x)$ is a lattice expression for the topological charge
density at gradient-flow time $t>0$. In the tests conducted here,
the so-called symmetric (``clover'') expression
was used for the charge density
and $t$ was set to the reference flow time $t_0$.

On large lattices, the susceptibility may be calculated by
splitting the sum over $y$ in two parts,
\begin{equation}
  \chi_t
  =
  \sum_{|y|\leq R}
  \langle q(y)q(0)\rangle+\delta(R)
  =
  \sum_{|y|\leq R}
  \llangle q(y)q(0)\rrangle+\delta(R)+\rmO(V^{-1/2}),
  \label{chit2}
\end{equation}
where
\begin{equation}
  \delta(R)=\sum_{|y|>R}
  \langle q(y)q(0)\rangle
  \label{chit3}
\end{equation}
is rapidly decaying at large $R$.
The replacement of the field-theoretical
correlation function in Eq.~(\ref{chit2})
by the translation average of the observable
\begin{equation}
  \obs(x)=\sum_{|y|\leq R}q(x+y)q(x)
  \label{chit4}
\end{equation}
adds a statistical error of order $V^{-1/2}$ to the expression.
A calculation of the susceptibility along these lines assumes that
there exist values of $R$,
where the systematic error $\delta(R)$
is small and which are much smaller than the linear extent of
the lattice so that the statistical error is small too.

Master-field simulations are, by definition, simulations where
the topological charge of the gauge field assumes some fixed value.
Fixed-topology simulations are known
to give results for local correlation functions
that differ from their exact field-theoretical values by terms
of order $V^{-1}$ \cite{BrowerEtAl,AokiEtAl}.
These effects are parametrically smaller than
the statistical errors, which shows that
master-field simulations provide a solution to the infamous
topology-freezing problem in lattice QCD if
the physical size of the lattice is large enough.

\begin{figure}[t] 
  \centering
  \includegraphics[width=6.8cm,clip]{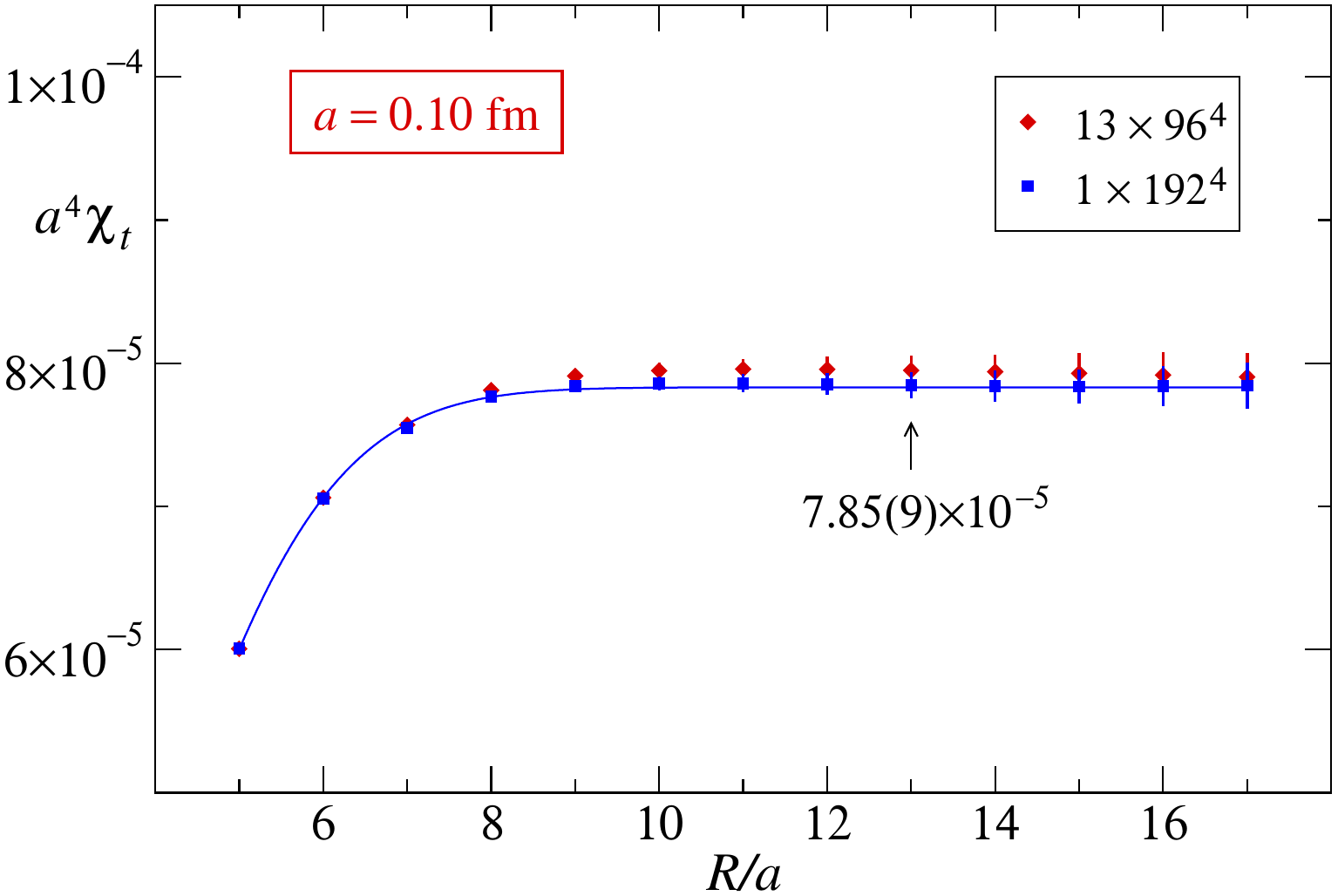}%
  \hspace{0.5cm}%
  \includegraphics[width=6.8cm,clip]{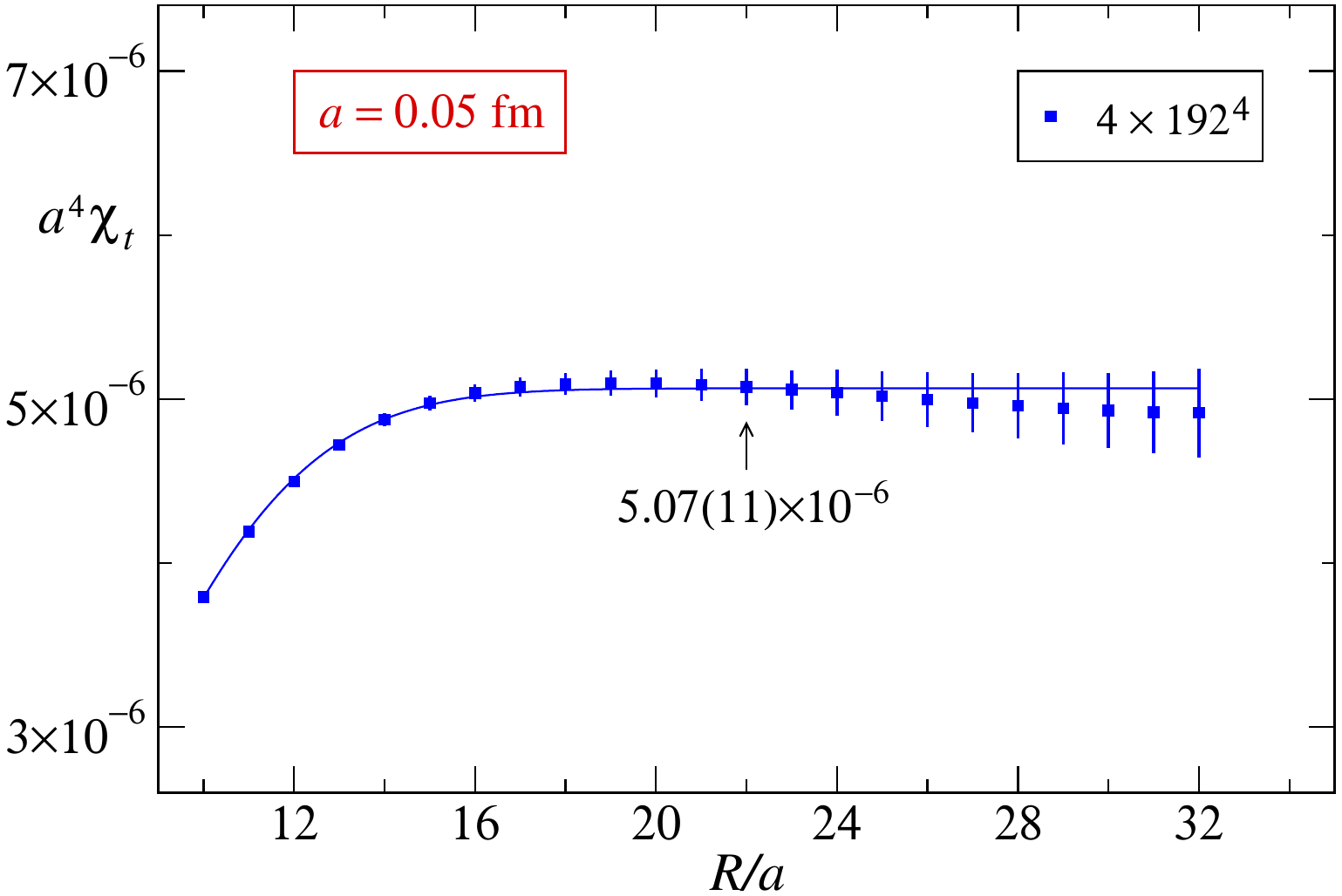}
  \caption{Master-field estimates of the topological susceptibility
           through the translation average of the observable (\ref{chit4})
           at two values of the lattice spacing. The smooth curves
           represent a fit of the data by a constant plus a
           deviation of the form (\ref{deltaR}).
           In these calculations,
           the systematic errors (\ref{chit3}) are estimated to be
           negligible with respect to the statistical errors at
           $R/a\geq13$ and $R/a\geq22$, respectively.
  }
  \label{fig3}
\end{figure}

As a function of the summation radius $R$,
the first term on the
far right of Eq.~(\ref{chit2}) reaches a plateau
at about $0.8$ fm (see Fig.~\ref{fig3}). Reflection positivity implies that
the second term $\delta(R)$ is negative at asymptotically
large values of $R$, but its analytic form is not known
and certainly strongly influenced by the gradient-flow smoothing
in the range of $R$,
where the data in Fig.~\ref{fig3} significantly depart
from the plateau value. The deviation
is actually well represented by
\begin{equation}
  \delta(R)=b\int_{|y|\geq R}\rmd^4y\,\rme^{-cy^2}
  \label{deltaR}
\end{equation}
in that range, with fitted parameters $b$ and $c$.

The results for the topological susceptibility in lattice units
quoted in Fig.~\ref{fig3} coincide, within errors, with
the values $7.884(9)\times10^{-5}$ \cite{CeEtAlTop}
and $4.98(30)\times10^{-6}$ \cite{ChowdhuryEtAl}%
\footnote{The value of the susceptibility quoted here is the
weighted average of the results obtained
on lattices with open and periodic boundary conditions
in time.}
previously obtained at $a=0.10$ and
$0.05$ fm, respectively, using traditional methods. At $a=0.10$ fm
the results are very precise and their agreement
is thus particularly impressive.
It might however be difficult to reach
a similar precision at $a=0.05$ fm with traditional simulations
in view of the long autocorrelation times
of the topological charge.
Master-field
simulations effectively bypass the problem and
yield results with constant relative precision as $a\to0$, if
the lattice volume is held fixed in physical units.

\subsection{Correlation function of the action density}\label{subsect33}

The calculation of the connected correlation function of the
Yang--Mills action density $E(x)$ in master-field simulations is,
in principle, straightforward. As is the case in ordinary
simulations, the signal-to-noise ratio decreases exponentially at
large distances and projections to a definite spatial momentum
further increase the statistical noise proportionally to the square
root of the three-dimensional volume (see ref.~\cite{KehFei} for
a recent discussion of the problem).
The increase of the statistical noise through the momentum projection
is however less severe in the case of
hadron propagators without disconnected parts. Here the issue is
avoided by considering
the point-to-point correlation function of the action density.

\begin{figure}[t] 
  \centering
  \includegraphics[width=6.0cm,clip]{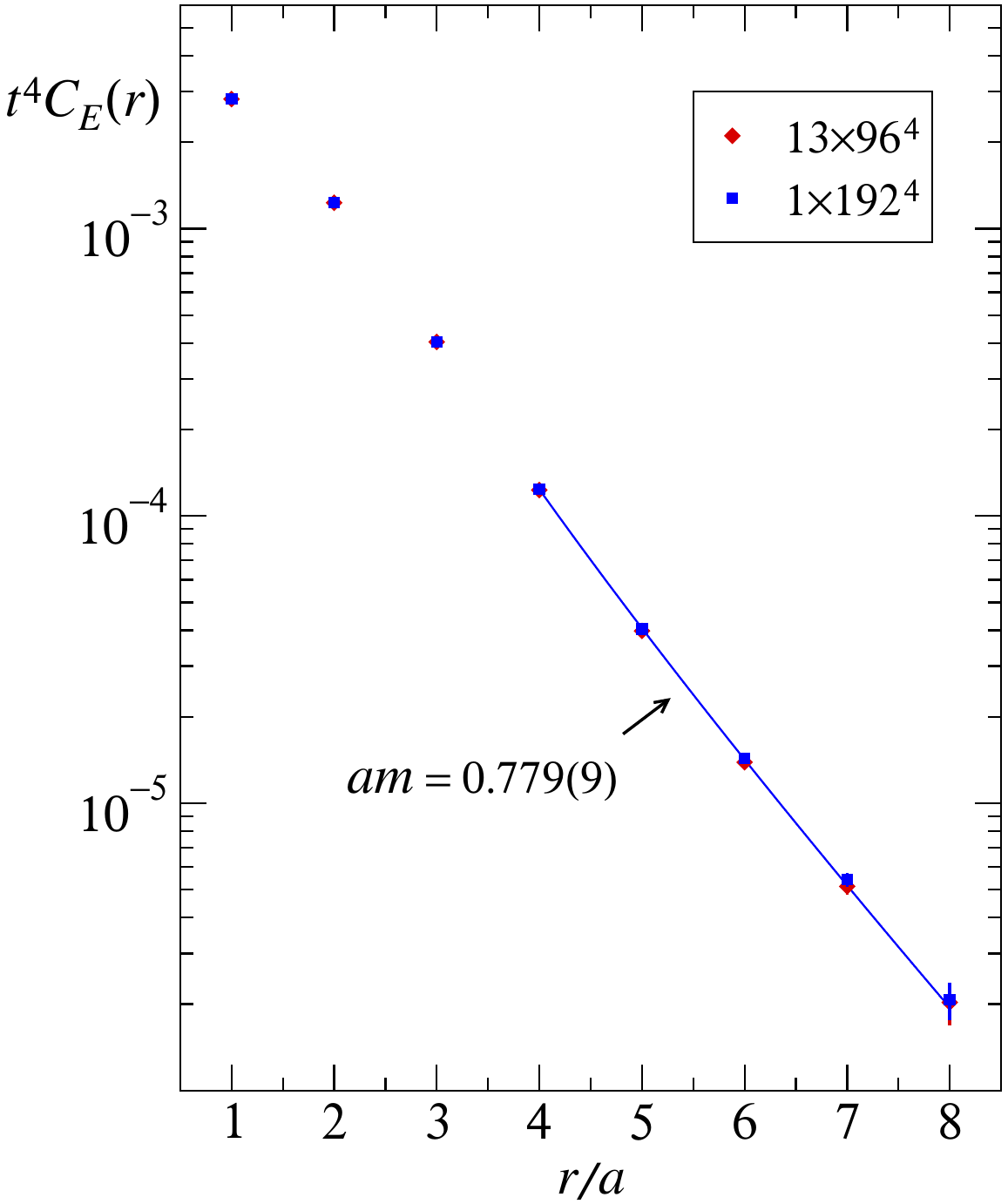}%
  \caption{Master-field simulation results for the
  correlation function (\ref{defCE}) of the action density
  at $a=0.10$ fm and gradient-flow time $t=0.36\times t_0$.
  The points obtained on the $96^4$ and $192^4$ lattices lie
  practically on top of each other. The full line represents
  an uncorrelated least-squares fit of the data on the larger lattice
  to the expected asymptotic form (\ref{asmCE}) of the correlation function.
  In the fit range, the statistical errors of $t^4C_E(r)$ determined by
  Eq.~(\ref{errCE2}) are roughly constant and equal to
  $3.1\times10^{-7}$ at $r/a=8$.
  }
  \label{fig4}
\end{figure}

Since the latter includes
a non-zero disconnected contribution,
the calculation requires two primary observables,
$\obs_1(x)$ and $\obs_2(x)$, to be chosen,
one for the connected and the other for the disconnected part.
In the present context, the observables should have a minimal
localization range centered at $x$. Moreover, assuming an
$L^4$ lattice, one would like to take advantage
of the hyper-cubic symmetry of the lattice theory to
reduce the statistical errors.

Considering these requirements,
a possible choice of observables is
\begin{equation}
  \obs_1(x)=\sfrac{1}{4}\sum_{\mu=0}^3
  E(x+h_{+}\hat{\mu})E(x-h_{-}\hat{\mu}),
  \quad
  \obs_2(x)=\sfrac{1}{8}\sum_{\mu=0}^3
  \bigl\{E(x+h_{+}\hat{\mu})+E(x-h_{-}\hat{\mu})\bigr\},
  \label{obsEE}
\end{equation}
where $\hat{\mu}$ denotes the unit vector in direction $\mu$ and
the displacements $h_{\pm}$ are integer multiples of the lattice
spacing. The (on-axis) correlation function at distance $r=h_{+}+h_{-}$
is then given by
\begin{equation}
  C_E(r)=\llangle\obs_1(x)\rrangle-
  \llangle\obs_2(x)\rrangle^2
  \label{defCE}
\end{equation}
up to statistical errors of order $V^{-1/2}$. In order to minimize
the localization region of the observables, $h_{+}$ is set to $r/2$ if
$r$ is an even multiple of the lattice spacing and to $(r+a)/2$
otherwise.

The square of the associated statistical error,
\begin{equation}
  \sigma_E(r)^2=
  \sum_{k,l=1}^2 w_kw_l\,\hbox{cov}_{kl}(r),
  \quad w_1=1,\quad w_2=-2\llangle\obs_2(x)\rrangle,
  \label{errCE}
\end{equation}
is determined by the standard error-propagation rules
and the $2\times2$ covariance matrix
\begin{equation}
  \hbox{cov}_{kl}(r)=\frac{1}{V}
  \sum_{|y|\leq R}\llangle\hat{\obs}_k(y)\hat{\obs}_l(0)\rrangle,
  \quad
  \hat{\obs}(x)\equiv\obs(x)-\llangle\obs(x)\rrangle,
  \label{varCE}
\end{equation}
(with a properly chosen value of the summation radius $R$)
of the primary observables.
Insertion of Eq.~(\ref{varCE}) in Eq.~(\ref{errCE}) followed by
a few lines of algebra then lead to the error formula
\begin{equation}
  \sigma_E(r)^2=
  \frac{1}{V}\sum_{|y|\leq R}
  \llangle \tilde{\obs}_1(y)\tilde{\obs}_1(0)\rrangle_{\rm c},
  \quad
  \tilde{\obs}_1(x)\equiv
  \sfrac{1}{4}\sum_{\mu=0}^3
  \hat{E}(x+h_{+}\hat{\mu})\hat{E}(x-h_{-}\hat{\mu}),
  \label{errCE2}
\end{equation}
which is easy to memorize given that Eq.~(\ref{defCE}) may be written
in the form
\begin{equation}
  C_E(r)=\llangle\tilde{\obs}_1(x)\rrangle.
  \label{defCE2}
\end{equation}
In the case
of several master fields, the same formulae hold with $\obs_1(x)$
and $\obs_2(x)$ replaced
by their ensemble averages (cf.~subsect.~\ref{subsect22}).

At large distances, the correlation function decays exponentially,
\begin{equation}
  C_E(r)\mathrel{\mathop\propto_{r\to\infty}}
  r^{-3/2}\rme^{-mr},
  \label{asmCE}
\end{equation}
where $m$ is the mass of the lightest scalar glueball.
The data shown in Fig.~\ref{fig4} can actually be accurately
represented by this asymptotic expression at distances $r\geq0.4$ fm.
Clearly, more
reliable determinations of the scalar glueball mass
would require the master-field simulations to be combined with
other techniques such as the variational method and multilevel
measurement algorithms.

\section{Generation of master fields}\label{sect4}

The derivation of the error formulae in sect.~\ref{sect2} shows
that any correct simulation algorithm may in principle
be used for the generation of master fields. If only
a single field is generated, all the computational work is done in
what is often referred to as the ``thermalization phase''
in ordinary simulations. Several fields are generated as usual
by running the algorithm for some length of time after
thermalization.
There are, however, important technical issues that must be addressed,
some concerning the thermalization phase and others the use of
global operations, which tends to be problematic on very large lattices.

\subsection{Thermalization}

\subsubsection{Building configurations from small lattices}

In order to reduce the computer time spent in the
thermalization phase, it is common practice to
construct the initial fields from thermalized fields
on smaller lattices through periodic extension.
Lattice extensions may proceed in several steps,
from one lattice size to the next
with intermediate thermalization runs, if initial fields
for simulations of very large lattices are constructed.

The topological charge of the fields generated in this way
however tends to grow proportionally to the lattice volume $V$,
particularly so when the autocorrelation time of the charge is
larger than the length of the intermediate thermalization runs.
For fixed-topology effects to be of order $V^{-1}$,
the value of the charge must be in the central range of the
exact charge distribution \cite{BrowerEtAl,AokiEtAl}.
Since the width of the distribution is of order $V^{1/2}$,
this condition is then likely to be violated.

The problem can easily be avoided by extending the fields through reflections
at the lattice planes rather than periodically. In this case, a
duplication of the field in one or more dimensions, for example,
produces a field with vanishing topological charge.
One then ends up with master fields that satisfy
the condition mentioned above with high probability.

\subsubsection{Autocorrelations}

To be able to judge whether the simulation has
reached its asymptotic stationary state,
the autocorrelation times of a representative set of
observables must be known.
Very accurate calculations of the autocorrelation times are not
required and the possibly very large autocorrelation times of the
topological charge can be ignored, since master-field simulations
are anyway fixed-topology simulations.
A sensible measure of autocorrelations,
which may be used in this context,
are the distances in simulation time, where
the autocorrelation functions of the observables considered have
decreased by, say, a factor $2$. These are easier to determine
than the integrated autocorrelation times and less sensitive
to any long tails of the autocorrelation
functions caused by the freezing of the topological charge.

If a simulation algorithm is used that updates the field variables
in physically distant regions essentially independently,
as does the SMD algorithm discussed below,
the autocorrelation functions of local observables can be expected to
decay rapidly in space. Under these conditions,
it is plausible that the autocorrelation times of
observables with localization ranges up to a few times the
basic correlation lengths capture the dynamics of the
algorithm sufficiently well. Moreover,
in the large-volume regime of QCD, their autocorrelation times
should be weakly dependent on the lattice size (if
the parameters of the theory and the algorithm are held fixed).
They can, therefore, be estimated on lattices much smaller
than the ones considered for the master-field simulations,
where long simulations may be unaffordable.

\subsection{Global operations}

The use of global branch conditions
and other global constructions is unnatural in local theories
and should be reconsidered when very large lattices are simulated.
Their implementation may, in any case,
be increasingly challenging in large volumes for purely
technical reasons.

\subsubsection{Solver stopping criterion}

At present all widely used algorithms for the solution of
the lattice Dirac equation are Krylov space
solvers with various preconditioners. If $D$, $\eta$ and $\psi$
denote the lattice Dirac operator, source field and current
approximate solution of the
equation, the solver program exits when the bound
\begin{equation}
  \|\eta-D\psi\|_2\leq\rho\,\|\eta\|_2,
  \qquad
  \|\eta\|^2_2\equiv\sum_x\eta(x)^{\dagger}\eta(x),
  \label{Stop}
\end{equation}
is satisfied,
where the tolerance $\rho$ is some fixed
small number such as $10^{-12}$.

The stopping criterion (\ref{Stop}) alone does not guarantee that
the residues $(\eta-D\psi)(x)$ of the calculated solutions
are uniformly small. In lattice QCD simulations,
where the norm of the source fields grows like $V^{1/2}$
with the lattice volume,
large local inaccuracies of the solutions (and thus
of the quark forces)
are not excluded and might compromise the simulations.
In principle the growth of the norm of the sources
with the lattice size could be compensated by
reducing the tolerance $\rho$, but the loss
of significance in the
calculation of the residues sets a lower
limit (of about $10^{-14}$ if
the fields are stored with 64 bit precision)
on the tolerances that can be imposed in practice.

On very large lattices,
the norm in Eq.~(\ref{Stop}) should thus be replaced
by a uniform norm such as
\begin{equation}
  \|\eta\|^2_{\infty}=\max_x\bigl\{\eta(x)^{\dagger}\eta(x)\bigr\}
  \label{MaxNorm}
\end{equation}
and solvers for the Dirac equation
must then be developed, which converge to
uniformly accurate solutions.
Alternating Schwarz procedures, local deflation and other
algorithms that proceed locally are likely elements of such solvers,
but much work will no doubt be required to find the right
combinations of these methods.

\subsubsection{HMC accept-reject step}

Global decisions are also taken
in the accept-reject step of the HMC algorithm,
which corrects for the inexact numerical integration
of the molecular-dynamics equations \cite{HMC}.
More precisely, the fields obtained at the end of the
molecular-dynamics evolution are accepted
with probability
\begin{equation}
  P_{\rm acc}=\min\{1,\rme^{-\Delta H}\},
  \label{Pacc}
\end{equation}
where $\Delta H$ is the difference of the values of the Hamilton function
at the beginning and the end of the evolution.
If an integration scheme of order $p$ is used, $\Delta H$
roughly scales like
\begin{equation}
  \Delta H\propto\eps^pV^{1/2}
  \label{DeltaH}
\end{equation}
with the integration step size $\eps$ and the lattice volume $V$.
In order to preserve a reasonable acceptance rate,
the step size must therefore be reduced proportionally
to $V^{-1/2p}$ if the lattice volume is increased.
More worrisome is
the fact that there is a loss of significance proportional to $V$
when $\Delta H$ is calculated.
On large lattices, the point where $\Delta H$ is obtained with
barely any significant digits may then rapidly be reached if the fields are
stored with 64 bit precision.

Both problems can be avoided by localizing the algorithm, i.e.~by
updating the field variables in subvolumes rather than all field
variables at once. In the case of the pure gauge theory, for example,
the link variables in any rectangular block of lattice points
can be updated with the HMC algorithm, while keeping the other field
variables fixed. The localization of the simulation is
however highly non-trivial in presence of the sea quarks and it is only
recently that a viable solution to the problem was found
by C\`e, Giusti and Schaefer \cite{Multilevel}.
So far the localized algorithm
has been described for divisions of the lattice in (thick) time slices,
but it is quite clear that algorithms localized in four dimensions
can be constructed along the same lines.

\subsection{SMD algorithm w/o accept-reject step}

Simply dropping the accept-reject step is
another possibility worth considering.
The algorithm then becomes inexact and its long-time behaviour
thus needs to be understood as well as the effects of the integration
errors on the calculated expectation values.
In the following, the stochastic-molecular-dynamics (SMD) algorithm
\cite{Horowitz,JansenLiu} is discussed, which is closely
related to the HMC algorithm and similarly efficient \cite{OpenBC},
but more easily accessible to rigorous mathematical analysis.

\subsubsection{SMD algorithm}

The SMD algorithm largely coincides with the HMC algorithm.
An update cycle starts with a refreshment of the momentum
field $\pi$ and the pseudo-fermion fields $\phi$ (if any)
according to
\begin{align}
   \pi&\to c_1\pi+c_2\upsilon,
   \qquad
   c_1=\rme^{-\gamma\eps},\quad c_1^2+c_2^2=1,\nonumber\\[2.0ex]
   \phi&\to c_1\phi+c_2\chi,
   \label{Rotation}
\end{align}
where $\upsilon$ and $\chi$ are random momentum and pseudo-fermion
fields with the appropriate Gaussian distributions. The algorithm
has two parameters, the friction constant $\gamma$ and
the simulation step size $\eps$, which are both assumed to be positive.
In the second step of the update cycle, the molecular-dynamics
equations for the momentum and the gauge field are integrated
from the current simulation time $t$ to $t+\eps$.
The accept-reject step normally applied at the end of the
molecular-dynamics evolution
is omitted here and the algorithm instead proceeds with the next
update cycle.

The simulation step size $\eps$ is usually set to a fraction of unity
and $\gamma$ is taken to be such that the accumulated
refreshment of the fields practically amounts to a
complete refreshment after running the algorithm
for a few units of simulation time.
Any symplectic reversible integrator,
such as the highly efficient ones described in ref.~\cite{OMF}, may be used
for the numerical integration of the molecular-dynamics equations.

\subsubsection{Existence and uniqueness of a stationary state}

Stochastic processes like the SMD algorithm need not
converge to a unique stationary state. They can run away in non-compact
directions of the field space, for example, or their long-time behaviour
may depend on the initial distribution of the fields.
Moreover, in the case of the SMD algorithm,
there is no explicitly known candidate for the
equilibrium distribution if the accept-reject step is dropped.

The theory of Markov processes on field spaces (and more general spaces)
is a well developed and active
research area in mathematics. In particular,
there exists a form of Harris' convergence theorem,
established by Hairer and Mattingly \cite{HairerMattingly},
which may be applied in the present context. Some work is
however still required to show that the premises of
the theorem are fulfilled \cite{SMDergodicity}.

As a result one obtains a rigorous proof of the convergence of
the SMD algorithm to a unique stationary state at large simulation
times, assuming the following properties hold:

\renewcommand{\labelitemi}{\textendash}

\begin{itemize}
\vspace{0.1cm}
\item{The gauge action is a smooth function of the link variables.}
\item{Sea quarks (if any) are included in the standard way, with
      operators in the pseudo-fermion actions that are bounded,
      strictly positive and smoothly dependent on the gauge
      field\,\footnote{In Wilson's formulation of lattice QCD,
      this condition can be met by including a small twisted-mass
      term in the Dirac operator as
      in ref.~\cite{openQCD}, for example.}.}
\item{A symplectic integrator of the separated type is
      used for the molecular-dynamics equations, where updates of
      the momentum and the gauge field alternate and the step sizes
      are proportional to the simulation step size $\eps$.}
\end{itemize}

\noindent
Convergence is then guaranteed if $\eps\leq\epsm$, where
$\epsm>0$ depends on the chosen gauge action and the molecular-dynamics
integrator, but not on the pseudo-fermion actions.
While the theorem does not exclude astronomically large autocorrelation times,
it asserts that the SMD algorithm without accept-reject step
eventually simulates a well-defined (albeit not analytically known)
probability distribution.

\subsubsection{Empirical studies}

Whether the SMD algorithm without accept-reject step
is a viable choice in practice also depends
on the size of the systematic errors that derive from the inexact
integration of the molecular-dynamics equations. In order to gain
some insight into this issue, the pure gauge theory was simulated
on a $64^4$ lattice with spacing $a=0.05$ fm and periodic boundary
conditions. The Wilson gauge action was used in this study and
the molecular-dynamics equations were integrated with a highly
efficient 4th-order scheme proposed by Omelyan et al.~\cite{OMF}
[the one with step sizes given by their Eqs.~(63) and (71)]. Following
ref.~\cite{OpenBC}, the friction parameter $\gamma$ was set to $0.3$
in all runs.

With these choices, the rigorous bounds obtained in the course
of the proof of the convergence theorem yield $\epsm\geq0.06\times g_0^2$,
where $g_0$ denotes the bare gauge coupling
($g_0^2\simeq0.935$ on the simulated lattices). Rigorous estimates
are usually far too pessimistic and up to at least $\eps=0.34$
no sign of an instability or slow convergence has
ever been observed.

\begin{figure}[t] 
  \centering
  \includegraphics[width=6.0cm,clip]{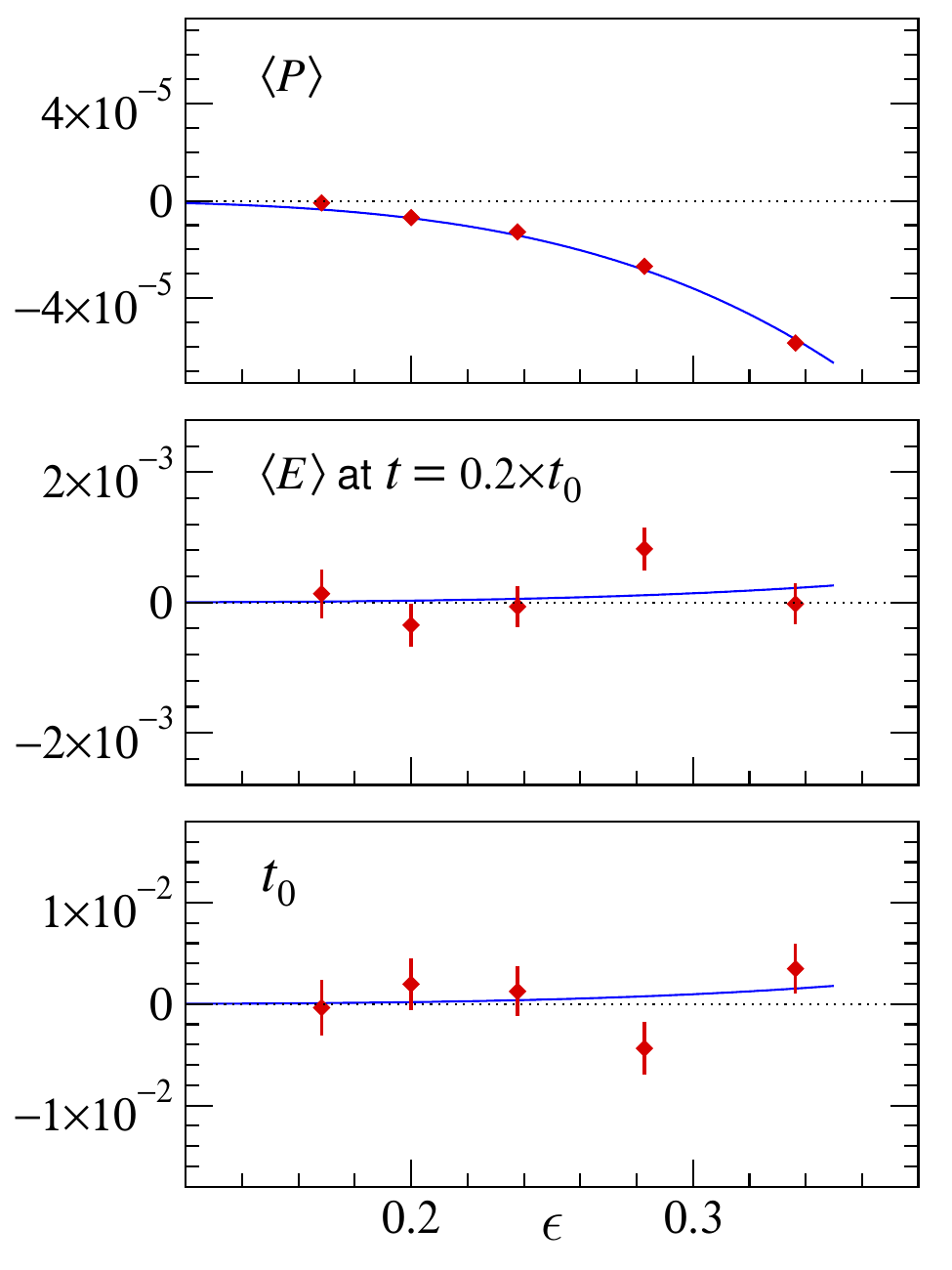}%
  \caption{Relative deviation from their exact values of
  the expectation value of the plaquette,
  the expectation value of
  the action density at a physically small flow time $t$
  and the reference flow time $t_0$
  as a function of the simulation step size $\eps$.
  The full lines represent fits of the data by a
  constant times $\eps^4$. No significant dependence on $\eps$ of
  the autocorrelation times was observed in these runs and the
  lengths of the simulations were about
  $1.8\times10^4$ units of molecular-dynamics time in all cases.
  }
  \label{fig5}
\end{figure}

Since a 4th-order integrator for the molecular-dynamics equations
was chosen, the effects of the integration errors on the
simulation results are expected to be of order $\eps^4$.
The data plotted in Fig.~\ref{fig5} suggest that they
are very small and thus difficult to observe.
Clearly visible effects are only seen in the case
of the plaquette expectation value, while the statistical errors of
the other quantities considered appear to be larger
than the integration errors.
Quantities sensitive to
the ultraviolet fluctuations of the gauge field are
actually likely to be the ones most strongly affected by
the inexact integration of the molecular-dynamics equations,
as is the case at leading order of perturbation
theory \cite{NSPT}.

The simulations reported in
sect.~\ref{sect3} were all performed using the SMD algorithm
as described here with simulation step size $\eps=0.1$.
This value of the step size is in the range of step sizes
one would use in HMC simulations, while the effects of
the integration errors are safely
negligible with respect to the quoted statistical errors.
The SMD algorithm without accept-reject step thus proves
to be a viable choice in the pure gauge theory, but the question will
obviously have to be addressed again when the sea quarks
are included in the simulations.

\section{Calculation of hadron propagators}\label{sect5}

The calculation of translation averages of hadron propagators
requires the latter to
be computed at all possible source points or at least on a sublattice
of points (cf.~sect.~\ref{subsubsect232}). Since the
Dirac equation must be solved a number of times for each source point,
the total computational effort then scales like $V^2$ on large lattices.

However, as explained below, the quark propagators are not needed
at arbitrarily large distances and they may consequently be
obtained with an effort growing proportionally to $V$
rather than $V^2$.
For illustration some algorithmic ideas are presented in this section,
but there could very well be better ones and they should, in any case,
be combined with variance reduction methods such as the
one described in ref.~\cite{Multilevel}.

\subsection{Basic strategy}

The principal idea is to exploit the fact
that the range of distances of interest normally extends
up to at most a few fm
and that the quark propagators fall off rapidly at large
distances.

Beyond the short distance regime, where the falloff is power-like,
the decay of the light-quark propagators is roughly described by
\begin{equation}
  |S(x,y)|\propto \exp\bigl\{-\sfrac{1}{2}\mpi|x-y|\bigr\},
  \label{Qprop}
\end{equation}
where $\mpi$ denotes
the mass of the pion at the chosen values of the bare parameters.
There is ample empirical evidence that Eq.~(\ref{Qprop}) holds
gauge configuration by gauge configuration on physically large lattices.
Differentiation of the propagator with respect to the gauge field
then shows that the sensitivity of the propagator on the gauge field
at lattice points far away from $x$ and $y$ is exponentially suppressed.

Up to small corrections,
one therefore expects to be able to compute the quark propagators
$S(x,y)$
by solving the Dirac equation in subvolumes of the lattice
containing $x$ and $y$.
The required shape and size of the subvolumes depends on the
desired level of accuracy and the range of distances considered.
In total the computational effort then scales proportionally
the number of source points times the size of the subvolumes.

\subsection{Domain decomposition}

\begin{figure}[t] 
  \centering
  \includegraphics[width=8.0cm,clip]{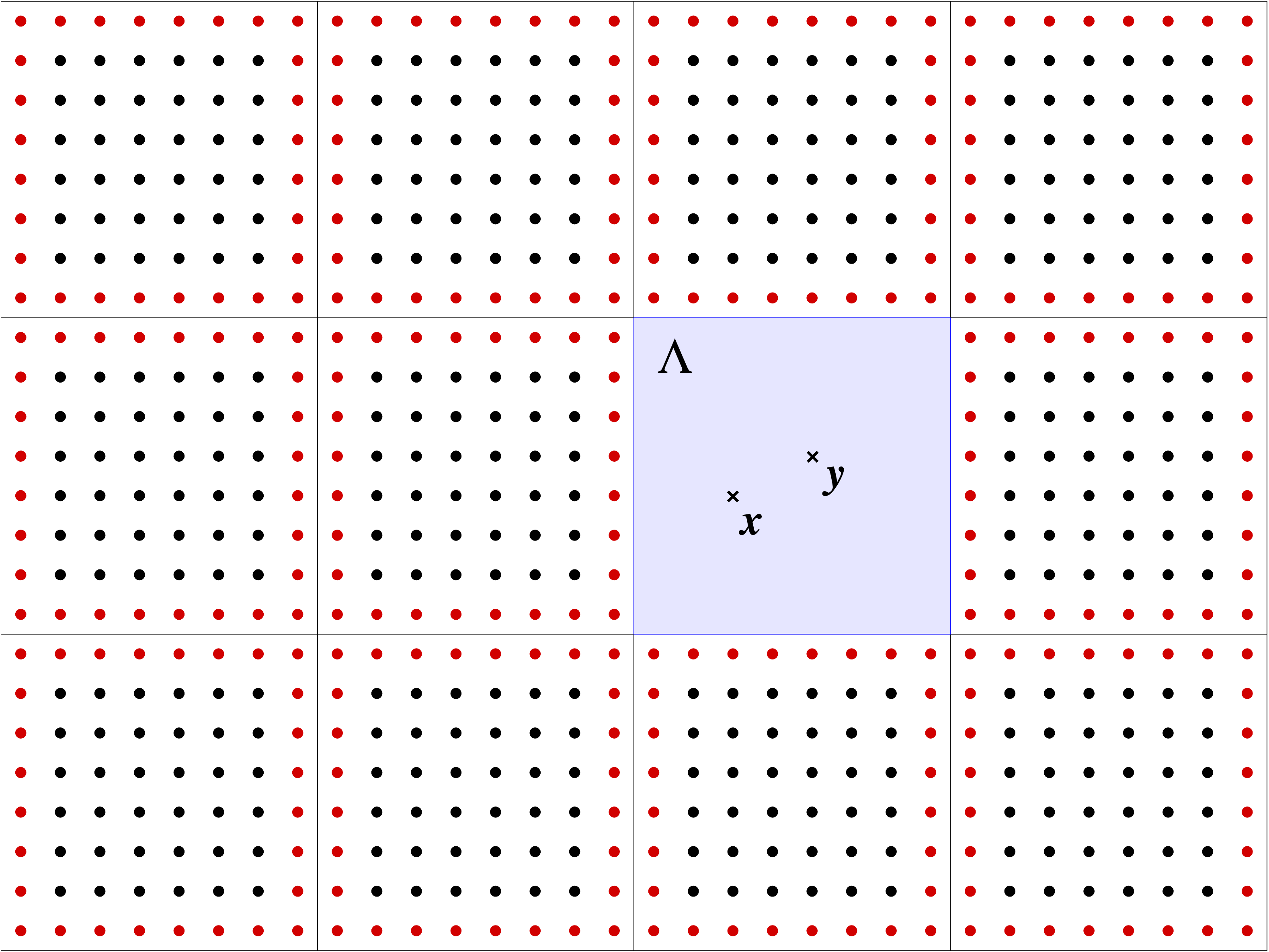}%
  \caption{Example of a division of the lattice into disjoint rectangular
  blocks $\Lambda$ of lattice points.
  The subset $\partial\Lambda^{\ast}$ of points in a block $\Lambda$
  with minimal distance from its boundaries (solid black lines)
  is referred to as the interior boundary
  of the block. In practice the blocks are taken to be several fm wide
  and normally contain many more points than shown in this sketch.
  }
  \label{fig6}
\end{figure}

In practice a possible way to proceed starts from a division of
the lattice into physically large blocks
as in Fig.~\ref{fig6}.
Such blocks of lattice points may be regarded as lattices on their own
with stochastic boundary conditions determined by the fields on
the surrounding lattice.

In the following, $D$ denotes the $\rmO(a)$-improved
Wilson--Dirac operator and $S(x,y)$ the associated quark propagator
on the full lattice in presence of a representative gauge field.
The renormalized quark mass is assumed be positive at the chosen
point in parameter space and the operator may include a twisted mass
term to safely exclude accidental zero modes.
The goal is then to accurately compute the propagators
at source points $y$ near the center of the
blocks and all $x$ at distances $|x-y|$
significantly smaller than the block edges.

On any given block $\Lambda$, let
\begin{equation}
  \Db=\Pb D\Pb,
  \qquad
  \Pb\psi(x)=
  \begin{cases}
     \psi(x) & \hbox{if $x\in\Lambda$,}\\[0.5ex]
     0       & \hbox{otherwise,}\\
  \end{cases}
  \label{Dblock}
\end{equation}
be the restriction of the Dirac operator to the block.
When acting on block fields,
$\Db$ coincides with the Dirac operator on $\Lambda$ with
homogeneous Dirichlet boundary conditions.
In particular, the
work required for the computation of the Green function
$S_{\Lambda}(x,y)$ of $\Db$ at all points $x\in\Lambda$
and any given source point $y\in\Lambda$ scales proportionally to the
volume of $\Lambda$.

If $x$ and $y$ are not too close to the boundary of $\Lambda$,
the propagator $S(x,y)$ is
well approximated by the block propagator $S_{\Lambda}(x,y)$.
It is possible to show this by decomposing the
Wilson--Dirac operator according to
\begin{equation}
  D=D_b+D_h,\qquad
  D_b=\sum_{\Lambda}\Db,
  \qquad
  D_h=\sum_{\Lambda}\Pb D(1-\Pb),
  \label{Dgrid}
\end{equation}
where $D_h$ is the sum of all hopping terms that go from one block to
another. Clearly, $D_b$ is a sum of commuting operators and
its Green function coincides with $S_{\Lambda}(x,y)$ on the block
$\Lambda$.
The identity
\begin{equation}
  \frac{1}{D}-\frac{1}{D_b}=-
  \frac{1}{D_b}D_h\frac{1}{D}
  \label{Dbnd}
\end{equation}
then leads to the formula
\begin{equation}
  S(x,y)-S_{\Lambda}(x,y)=-
  \sum_{z\in\partial\Lambda^{\ast}}
  S_{\Lambda}(x,z)(D_hS)(z,y),
  \qquad x,y\in\Lambda,
  \label{Sbnd}
\end{equation}
for the propagator difference. Recalling Eq.~(\ref{Qprop}),
the difference is thus seen to be exponentially
sup\-pressed with respect to $S(x,y)$
if both $x$ and $y$ are well inside the block.

\subsection{Stochastic representation}

Let $\eta$
be a random pseudo-fermion field with normal distribution
on the interior boundaries $\partial\Lambda^{\ast}$ of
the blocks $\Lambda$ (the set of points
along the block boundaries in Fig.~\ref{fig6}).
Equation (\ref{Sbnd}) may then be written in the form
\begin{equation}
  S(x,y)=S_{\Lambda}(x,y)+
  \bigl\langle\phi_{\Lambda}(x)\otimes\chi(y)^{\dagger}\bigr\rangle_{\eta},
  \qquad
  \phi_{\Lambda}=-D_{\Lambda}^{-1}D_h\eta,
  \qquad
  \chi=(D^{\dagger})^{-1}\eta.
  \label{SR1}
\end{equation}
For each random field configuration and all points
$x,y$ well inside $\Lambda$,
the product of fields in the second term on the right of this formula
is exponentially suppressed.
A sample of a few random fields may therefore allow
the term to be estimated to sufficient precision,
in which case the quark propagators are obtained
with a computational effort that scales proportionally to $V$
times the number of source points per block,

\section{Conclusions}\label{sect6}

Master-field simulations extend the scope of numerical lattice QCD
and are likely to have many interesting applications.
They might sometimes be used in place of traditional simulations,
where smaller lattices are simulated through large ensembles of
representative fields, but whether this is profitable depends
very much on the observables considered and the physics questions
that are to be addressed. For computations of scattering phases
for example, very large volumes are
(somewhat paradoxically) not necessarily advantageous.

Apart from giving access to a new kinematic regime, master-field simulations
provide a solution to the topology-freezing problem, which tends to affect
lattice QCD simulations near the continuum limit.
There are, however, still a few technical questions
that must be addressed when the sea quarks are included
in master-field simulations.
In particular, global branch conditions tend to be
problematic on very large lattices and should be critically
reviewed. Methods based on domain
decompositions and multilevel algorithms,
on the other hand, are natural strategies on these lattices
and are anyway recommended
for reasons of algorithmic efficiency and ease of parallel processing.

In practice the memory requirements of master-field simulations
can be challenging.
On a $256^4$ lattice, for example, simulations of the SU(3)
gauge theory require about $5$ TB of memory, and this figure easily
rises to $100$ TB or more when the sea quarks
are included in the simulations. Alternative program structures are however
conceivable, where large parts of the
lattice are visited in order, while
the fields residing elsewhere
are preserved on some storage devices other
than the main memory.
For reading and writing gauge-field configurations,
the parallel I/O bandwidth offered by common
HPC infrastructures is currently not a bottleneck.

\begin{acknowledgement}

I wish to thank the organizers of this year's lattice conference
for the kind invitation to present this
work. Thanks also go to the University of Granada for hospitality at
the Carmen de la Victoria, its wonderful guest house.
While preparing this talk, I profited from
inspiring discussions with Leonardo Giusti, and
the first master-field simulations of some very large lattices
could not have been completed before the conference
without the help of Isabel Campos.

The simulations reported here were performed on a dedicated
HPC cluster at CERN and on the
FinisTerrae II machine provided by CESGA (Galicia Supercomputing Centre) to
IFCA-CSIC in the framework of the H2020 project INDIGO-Datacloud (RIA
653549).
FinisTerrae II was funded by the Xunta de Galicia and the Spanish MINECO under
the 2007-2013 Spanish ERDF Programme.
I am grateful to all these institutions
for the generous support given to this work.

\end{acknowledgement}

\end{document}